\documentclass{IEEEtran}
\usepackage{soul, color, xcolor}
\usepackage{cite}
\usepackage{amsmath,amssymb,amsfonts}
\usepackage{graphicx}
\usepackage{textcomp,nicefrac}
\def\BibTeX{{\rm B\kern-.05em{\sc i\kern-.025em b}\kern-.08em
T\kern-.1667em\lower.7ex\hbox{E}\kern-.125emX}}
\begin{document}
\title{Capacitive coupling study of the HERD SCD prototype: preliminary results }
\author{Ruo-Si Lu, Rui Qiao, Ke Gong, Wen-Xi Peng, Wei-Shuai Zhang, Dong-Ya Guo, Jia-Ju Wei, Yi-Ming Hu, Jian-Hua Guo, Qi Wu, Peng Hu, Xuan Liu, Bing Lu, Yi-Rong Zhang
\thanks{This research is supported by the States Key Project of Research and Development Plan (2021YFA0718403, 2021YFA0718404, 2022YFF0503302), the National Natural Science Foundation of China (Projects: 12061131007).}
\thanks{Ruo-Si Lu, Rui Qiao, Ke Gong, Wen-Xi Peng, Wei-Shuai Zhang, Dong-Ya Guo, Qi Wu, Peng Hu, Xuan Liu,  Bing Lu and Yi-Rong Zhang are with the Institute of High Energy Physics, Chinese Academy of Sciences, Beijing 100049, China (e-mail: lurs@ihep.ac.cn; qiaorui@ihep.ac.cn; gongk@ihep.ac.cn; pengwx@ihep.ac.cn; zhangws@ihep.ac.cn; guody@ihep.ac.cn; wuqi98@ihep.ac.cn; hupeng@ihep.ac.cn; liuxuan@ihep.ac.cn; lubing@ihep.ac.cn; 2021226055026@stu.scu.edu.cn).}
\thanks{Ruo-Si Lu, Rui Qiao, Ke Gong, Wen-Xi Peng and Qi Wu are with the University of Chinese Academy of Sciences, Beijing 100049, China (e-mail: lurs@ihep.ac.cn; qiaorui@ihep.ac.cn; gongk@ihep.ac.cn; pengwx@ihep.ac.cn; wuqi98@ihep.ac.cn).}
\thanks{Jia-Ju Wei, Yi-Ming Hu, and Jian-Hua Guo are with the Key Laboratory of Dark Matter and Space Astronomy, Purple Mountain Observatory, Chinese Academy of Sciences, Nanjing 210008, China (e-mail: weijj@pmo.ac.cn; huyiming@pmo.ac.cn;  jhguo@pmo.ac.cn).}
\thanks{Jian-Hua Guo is with School of Astronomy and Space Science, University of Science and Technology of China, Hefei 230026, China (e-mail: jhguo@pmo.ac.cn).}
\thanks{Wei-Shuai Zhang is with School of Physics and Astronomy, China West Normal University, Nanchong 637002, China (e-mail: zhangws@ihep.ac.cn).}
\thanks{Xuan Liu is with North China University of Technology, Beijing, 100144, China (e-mail: liuxuan@ihep.ac.cn).}
\thanks{Yi-Rong Zhang is with Sichuan University, Chengdu, 610065, China (e-mail: 2021226055026@stu.scu.edu.cn).}}

\maketitle

\begin{abstract}
The Silicon Charge Detector (SCD) is a subdetector of the High Energy Cosmic Radiation Detection payload. The dynamic range of the silicon microstrip detector can be extended by the capacitive coupling effect, which is related to the interstrip capacitance and the coupling capacitance. A detector prototype with several sets of parameters was designed and tested in the ion beams at the CERN Super Proton Synchrotron. The capacitive coupling fractions with readout strip and floating strip incidences were studied using the beam test data and SPICE simulation.
\end{abstract}


\section{Introduction}
\label{sec:introduction}
\IEEEPARstart{T}{he} High Energy Cosmic Radiation Detection (HERD) facility is a dedicated particle and astrophysical experiment for the Chinese space station. It aims to achieve several key scientific objectives, including indirect searches for dark matter, precise spectroscopy, and composition measurements of cosmic rays, and monitoring high-energy gamma rays [1]. One of the unresolved phenomena in cosmic ray detection is the ``knee", which refers to the steepening of primary cosmic rays near the PeV energy range [2]. The operation of HERD is expected to make significant contributions to understanding this phenomenon.

The HERD facility comprises a 3-D cubic imaging calorimeter (CALO) surrounded by five sides of trackers, Plastic Scintillator Detector (PSD) and Silicon Charge Detector (SCD) [3]. The envelope size of the HERD facility is 3.0 × 2.3 × 1.7 m$^{3}$. SCD is located at the outmost of HERD, as shown in Fig. 1.

The SCD is designed to measure the charge of high-energy cosmic ray nuclei ranging from Z of 1 to about 28. The SCD is composed of a Top-SCD unit covering an area of 1.8 × 1.8 m$^{2}$, and four Side-SCD units with an area of 1.6 × 1.1 m$^{2}$ each. Each SCD unit consists of eight layers of single-sided silicon microstrip detectors. The adjacent layers are installed in orthogonal directions to identify the charge and trajectories of incoming charged particles [4].

\begin{figure}[t]
\centerline{\includegraphics[width=3.5in]{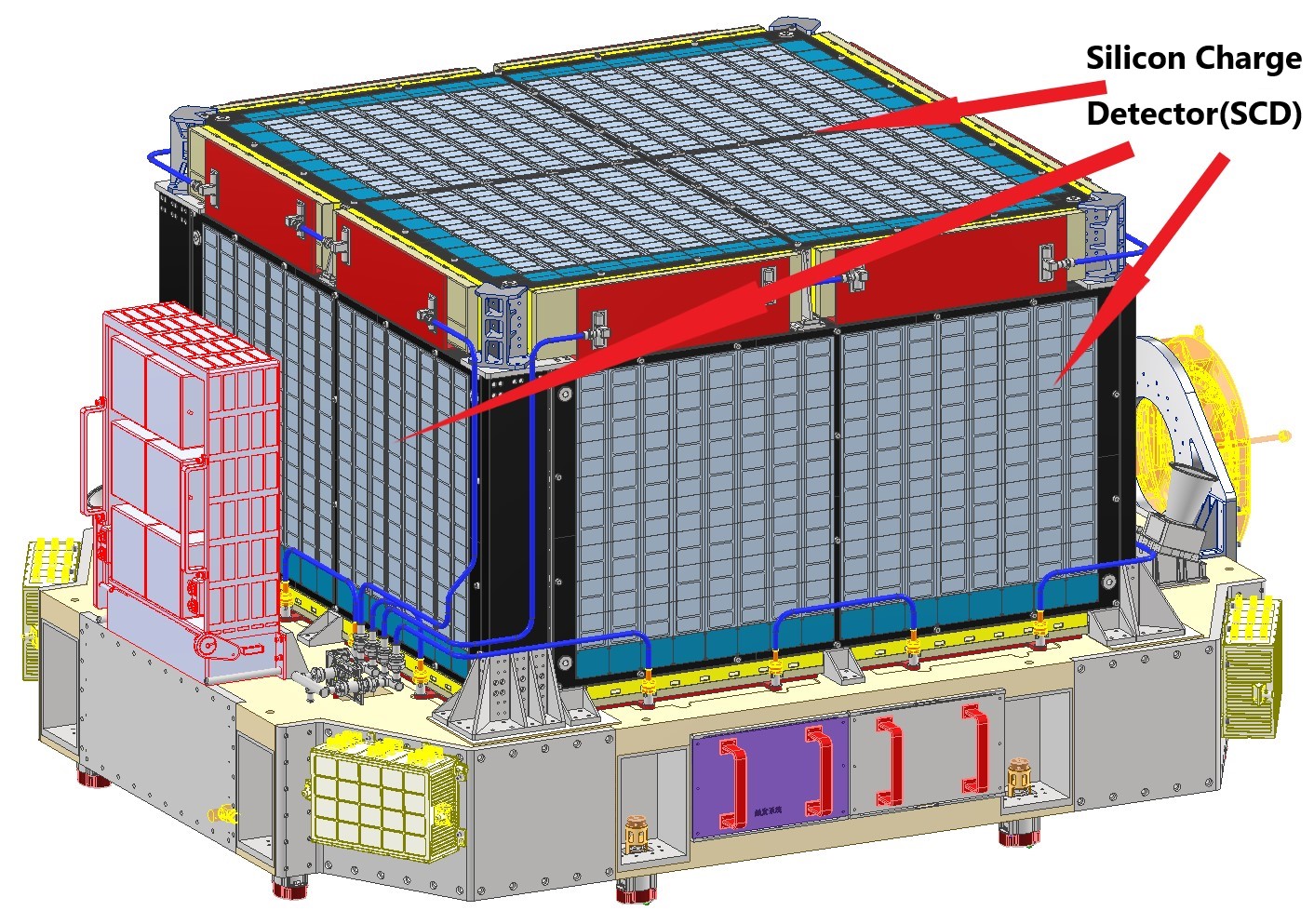}}
\caption{Schematics of the HERD detector.}
\label{fig1}
\end{figure}

A silicon microstrip detector can be modeled as a network of capacitors, which includes the bulk capacitors, the interstrip capacitors, and the coupling capacitors [5,6]. When a charge signal is generated within a strip that has been hit, it can be capacitively coupled to neighboring strips through the capacitor network. This capacitive coupling effect is negligible when the coupling capacitance is significantly larger than the interstrip capacitance. On the contrary, this capacitive coupling effect can be enhanced by using smaller coupling capacitors as discussed in this paper.

The electronic design of SCD is inherited from the Silicon Tungsten Tracker of the Dark Matter Explorer (DAMPE), whose linear dynamic range can only directly measure the signal of Z = 1 $\sim$ 6. The SCD is proposed to increase the dynamic range to measure the signal of Z = 1 $\sim$ 28 by enhancing the capacitive coupling effect. The small signals from low-Z particles can be easily measured using the fired strip. The large signals from high-Z particles could saturate the readout electronics of the fired strip, but the capacitively coupled signals of the neighboring strips are not saturated. The total signals can be reconstructed as the coupled signals divided by the coupling fractions. This helps the SCD to measure high-Z particles and extend the dynamic range. However, if the capacitive coupling fractions are too small (Fig. 2a), the coupled signals may be too weak and the signal-to-noise ratios are poor. The capacitive coupling fractions of SCD are preferred to be larger by approximately one order of magnitude than the fractions without external capacitors (Fig. 2b). The dependence of capacitive coupling fractions and the capacitor network of the SCD prototype detector was investigated through ion beam test and SPICE simulation.

\begin{figure}[t]
\centerline{\includegraphics[width=3.5in]{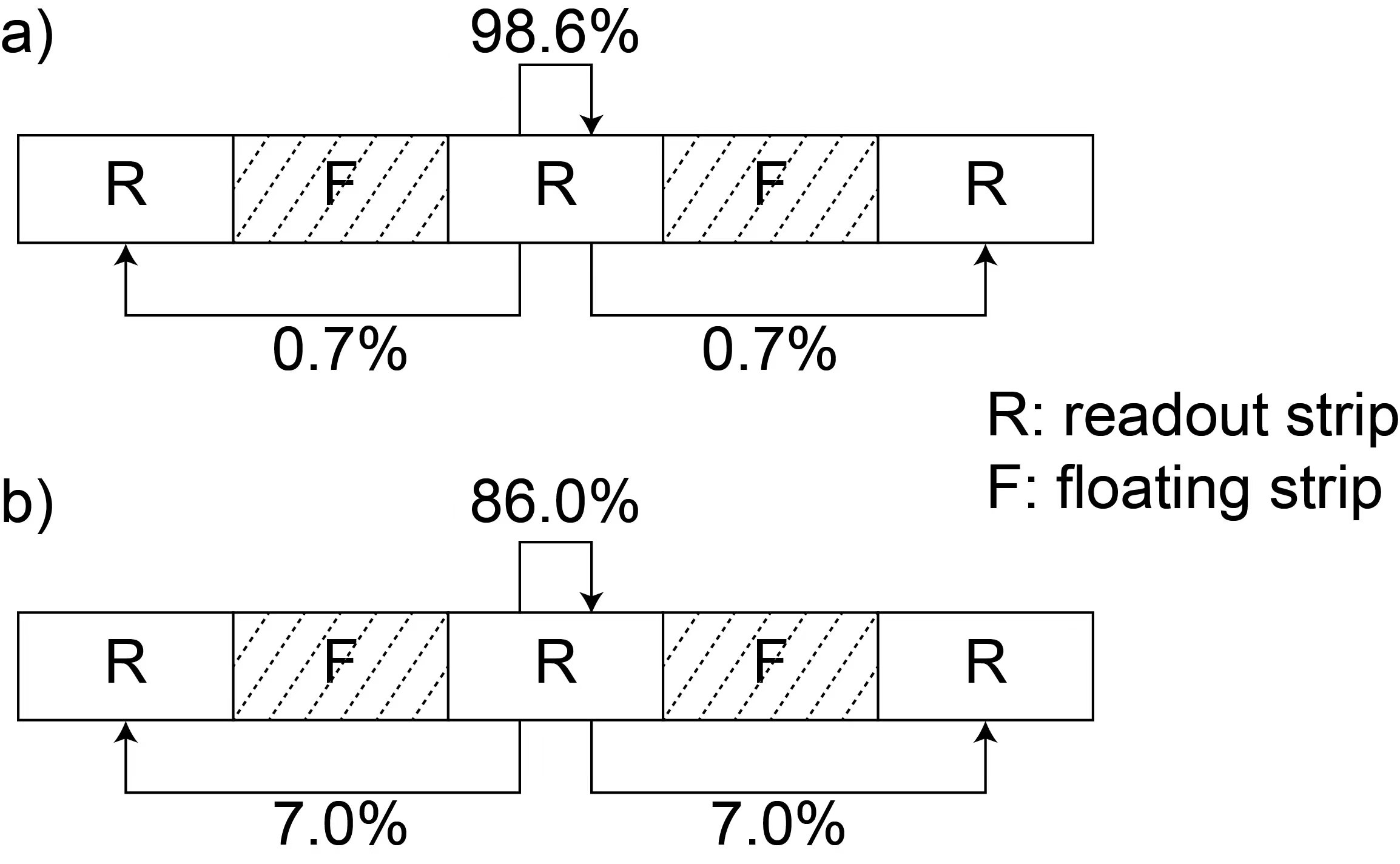}}
\caption{The capacitive coupling fractions of set 2 a) and set 4 b). The definition of the two sets will be discussed below}.
\label{fig2}
\end{figure}

\section{The SCD prototype detector}
The single-sided AC-coupled prototype detector (Products ID: OOO-2), ordered from MICRON semiconductor in 2021 [7], has a thickness of 300 $\mu$m and an active area of 6 cm × 3.2 cm (Fig. 3). The full depletion voltage of the sensor is 30 V and it was biased at 80 V during the experiment. Fig. 4 illustrates a cross-section of the silicon sensor. The junction side of the detector has 400 p+ strips. The implantation and readout pitch are 80 $\mu$m and 160 $\mu$m, respectively. The detector is divided into two groups: the first half with 200 strips have a width of 60 $\mu$m, while the remaining 200 strips have a width of 25 $\mu$m. Half of the implantation strips are AC-coupled to the front-end electronics and amplified by four 64-ch IDE1140 ASICs [8], while the other half implantation strips remain floating. These two types of implantation strips are referred to as the readout strips and floating strips, respectively. This interval readout design allows for achieving good spatial resolution using limited readout channels [9,10].

\begin{figure}[t]
\centerline{\includegraphics[width=3.5in]{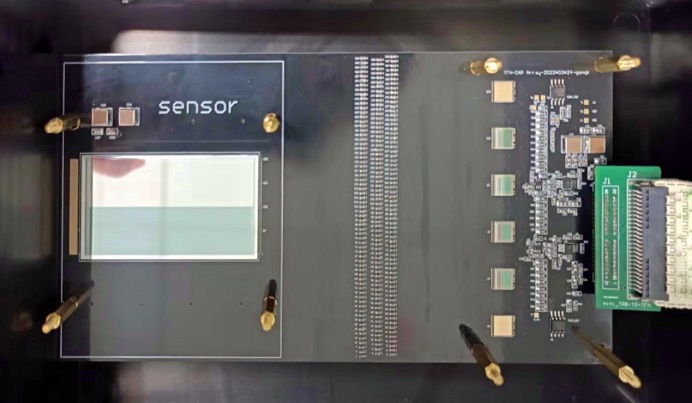}}
\caption{View of the SCD prototype detector.}
\label{fig3}
\end{figure}

In order to investigate the dependence of capacitive coupling on the coupling capacitors, the 100 readout channels in each group are divided into five sets (Fig. 5). Each set consists of 20 readout channels coupled to various external capacitors, as listed in Table 1.

\begin{figure}[t]
\centerline{\includegraphics[width=3.5in]{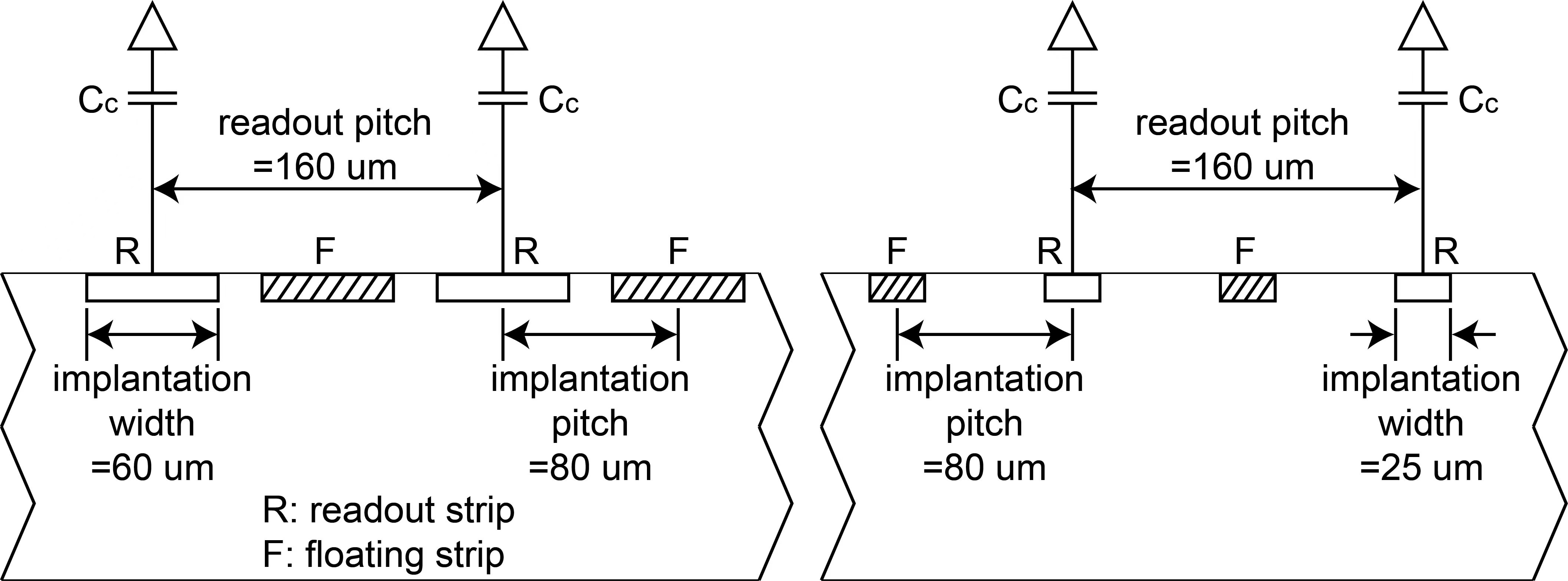}}
\caption{Layout of the silicon microstrip sensor.}
\label{fig4}
\end{figure}

\begin{figure}[t]
\centerline{\includegraphics[width=3.5in]{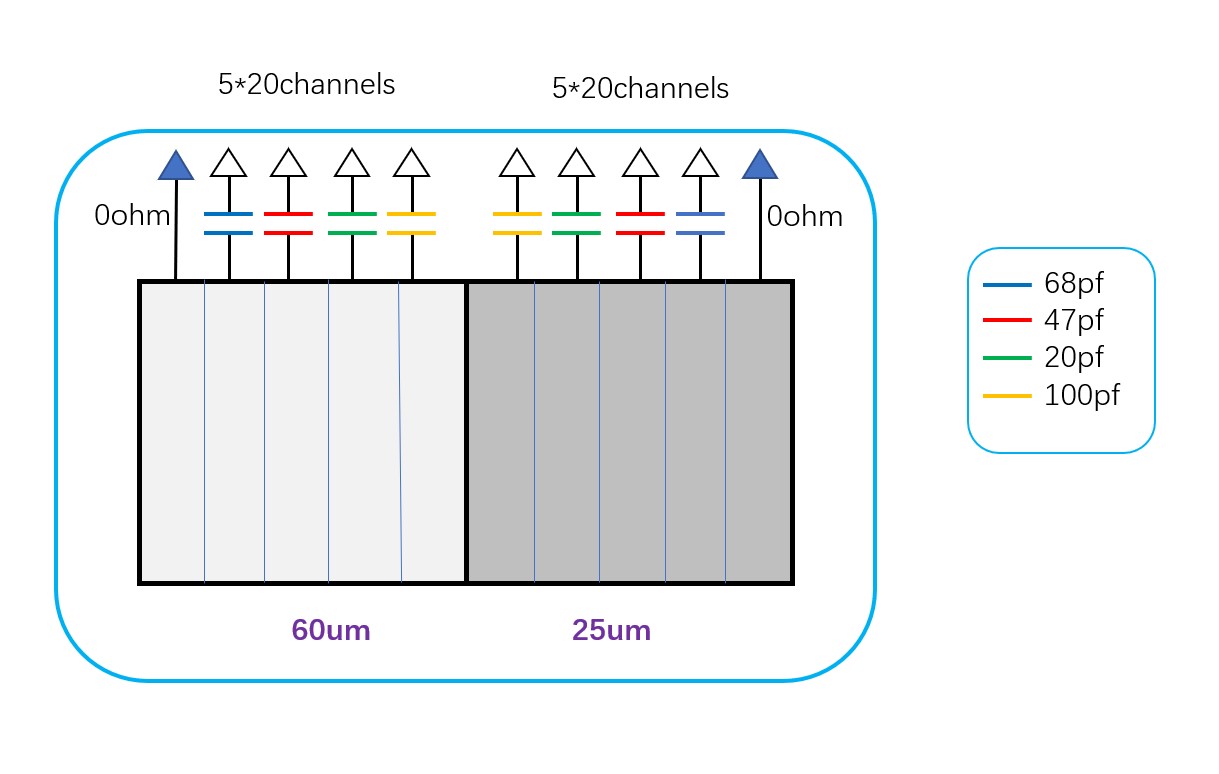}}
\caption{Schematic diagram of a detector with five sets for each of the two strip widths.}
\label{fig5}
\end{figure}

\begin{table}
\caption{Parameters for each set of the detector}
\label{table1}
\setlength{\tabcolsep}{3pt}
\begin{tabular}{|p{35pt}|p{75pt}|p{105pt}|}
\hline
Set&
Strip width ($\mu$m)&
External capacitors (pF) \\
\hline
1&
60&
68\\
2&
60&
N/A\\
3&
60&
47\\
4&
60&
20\\
5&
60&
100\\
6&
25&
100\\
7&
25&
20\\
8&
25&
47\\
9&
25&
N/A\\
10&
25&
68\\
\hline
\end{tabular}
\end{table}

The readout channels of both set 1 and set 10 (with 68 pF external capacitors) are connected to two different ASICs. This can introduce a bias in the capacitive coupling effect due to the different ASIC gains. Besides, strip width of 60 $\mu$m is closer to the final design. Consequently, only the analysis results of set 2 to set 5 are presented and discussed.

\section{Capacitance Measurement}
The bulk capacitance, coupling capacitance, and interstrip capacitance of the SCD were measured using an Agilent 4980A LCR meter. The detector was biased at 80 V using a Keithley 6487 picoammeter. An Agilent 16065A external voltage bias fixture was used to prevent the bias voltage from damaging the LCR meter. For each type of capacitance, the mean value and the error were calculated through five measurements. In addition, an Agilent 16380A standard capacitor set was used for calibration before each measurement.

The coupling capacitance (C\textsubscript{c}) was measured by probing the AC pad and DC pad of the same strip. The equivalent parallel capacitance (C\textsubscript{p}) and series capacitance (C\textsubscript{s}) were read from the LCR meter within a frequency range from 100 Hz to 2 MHz, as shown in Fig. 6. At low frequencies, C\textsubscript{p} and C\textsubscript{s} were the same and independent of the frequency. As the frequency increased, the resistive implantation strip gradually blocked the AC signal, causing a decrease in the measured capacitance [5]. The measurements at 600 Hz were chosen to evaluate the coupling capacitance.

\begin{figure}[t]
\centerline{\includegraphics[width=3.5in]{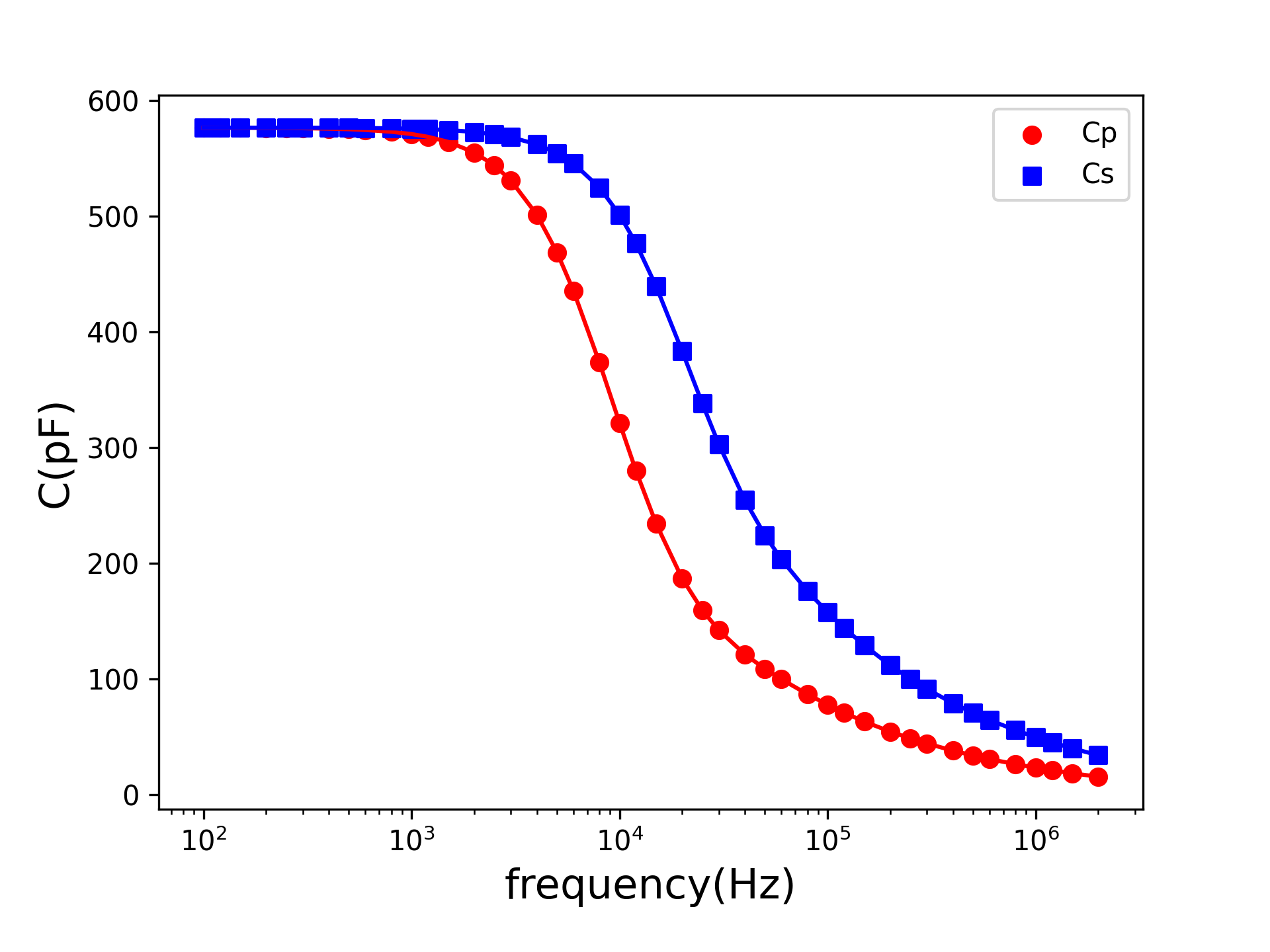}}
\caption{The coupling capacitance dependence on the frequency.}
\label{fig6}
\end{figure}

The total bulk capacitance was measured by probing the bias ring and the  backplane of the detector. The bias voltage was scanned from 5 V to 80 V at a frequency of 600 Hz, as shown in Fig. 7. Two linear fits were conducted, and the full depletion voltage was approximately 29.9 V. The bulk capacitance of a single strip (C\textsubscript{b}) is calculated by dividing the total bulk capacitance by the number of strips.

\begin{figure}[t]
\centerline{\includegraphics[width=3.5in]{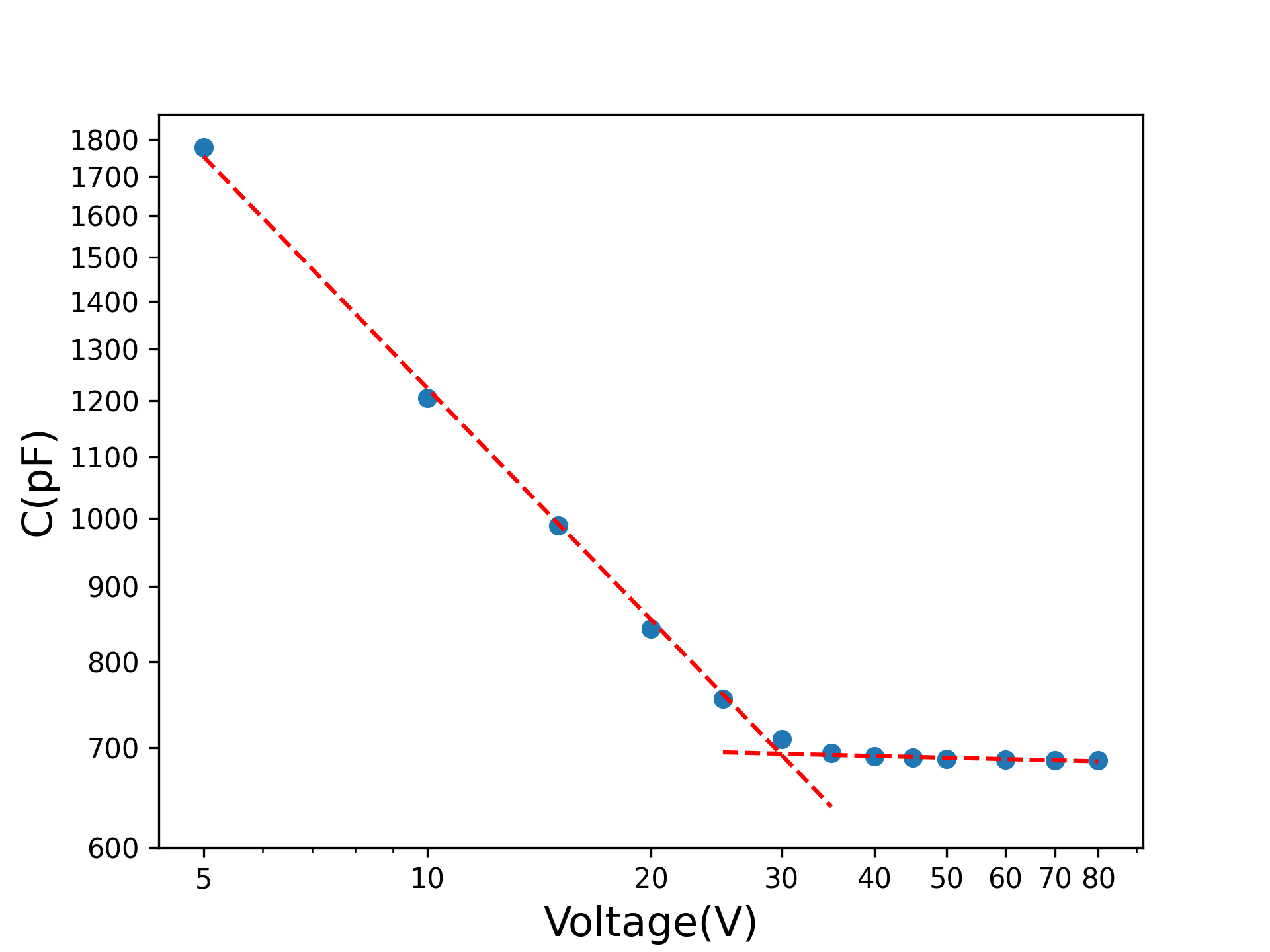}}
\caption{The total bulk capacitance as a function of the bias voltage.}
\label{fig7}
\end{figure}

Three methods were used to measure the first interstrip capacitance (C\textsubscript{i1}) by placing the probes on the pads of two adjacent strips: a) AC-AC pads, b) DC-DC pads and c) AC-DC pads. The measurements were conducted with 10 kHz where the capacitance was independent of the frequency. The capacitance obtained from the three methods were 6.18 ± 0.16 pF, 6.27 ± 0.27 pF, and 6.35 ± 0.27 pF, respectively. They agreed with each other within the range of error. Therefore, only the AC-AC pads were measured for the second interstrip capacitance (C\textsubscript{i2}) and third interstrip capacitance (C\textsubscript{i3}). The final results were shown in Table 2.

\begin{table}
\caption{The capacitance measurement results}
\label{table2}
\setlength{\tabcolsep}{3pt}
\begin{tabular}{|p{55pt}|p{75pt}|p{85pt}|}
\hline
Capacitance&
Value (pF)&
Error (pF)\\
\hline
Cc&
575&
0.65\\
Cb&
1.66&
N/A\\
Ci1&
6.27&
0.29\\
Ci2&
0.26&
0.30\\
Ci3&
0.21&
0.12\\
\hline
\end{tabular}
\end{table}

\section{Ion beam test results }
\subsection{Experimental setup}
An ion beam test was conducted at the CERN Super Proton Synchrotron (SPS) in October 2022 to investigate the response of ions for all the HERD prototype detectors. The test involved a 150 GeV/n lead primary beam directed at a 4 cm thick beryllium target. The secondary particles were selected using magnets and then directed towards the HERD prototype detectors. The SCD was mounted on a moving platform perpendicular to the ion beam, and the height of the platform was adjusted during the test to ensure that most of the detector sets received illumination from the small collimated beams.

\subsection{Raw data process and event selection}
The raw data processing consists of three steps: pedestal subtraction, common noise subtraction, and cluster finding. Firstly, the pedestal of each channel was determined by calculating the average ADC value recorded during a pedestal run for that specific channel. This pedestal value was then subtracted from the channel amplitude in every beam run to remove the baseline. Secondly, the common noise, which is caused by power supply fluctuations, was calculated as the average ADC value of each ASIC and then subtracted event-by-event [11]. Lastly, a cluster finding algorithm was applied to identify all the clusters [11], and the maximum cluster is selected. The amplitudes of the maximum channel within the cluster (known as the seed channel) along with its eight neighboring channels were preserved for further analysis.
Events with readout strips or floating strips incidences were selected based on the impact position $\eta$, defined as: 
\begin{equation}
\eta =
\begin{cases}
\frac{PH_{\text{seed}}}{PH_{\text{seed}-1}+PH_{\text{seed}}} & \text{if } PH_{\text{seed}-1} \geq PH_{\text{seed}+1}\\
\frac{PH_{\text{seed}+1}}{PH_{\text{seed}}+PH_{\text{seed}+1}} & \text{if } PH_{\text{seed}-1} < PH_{\text{seed}+1}\
\end{cases}
\label{eq:eta}
\end{equation}

where $PH_{seed}$, $PH_{seed-1}$ and $PH_{seed+1}$ are the signal amplitudes of the seed channel and its two neighboring channels after pedestal and common noise subtractions, respectively.

When a particle hits a readout strip, the majority of the charge is collected by the seed channel, resulting in the $\eta$ value close to 0 or 1. On the other hand, when a particle hits the floating strip, the charge is capacitively coupled and distributed almost equally between the two adjacent readout strips, leading to the $\eta$ value close to 0.5. Fig. 8 illustrates the distribution of $\eta$ values for incident lithium nuclei on the detector set 4. The readout strip incidence is identified with $\eta$ values within the range of [0, 0.15] and [0.85, 1], while the floating strip incidence is identified with $\eta$ values between 0.45 and 0.55. 

\begin{figure}[t]
\centerline{\includegraphics[width=3.5in]{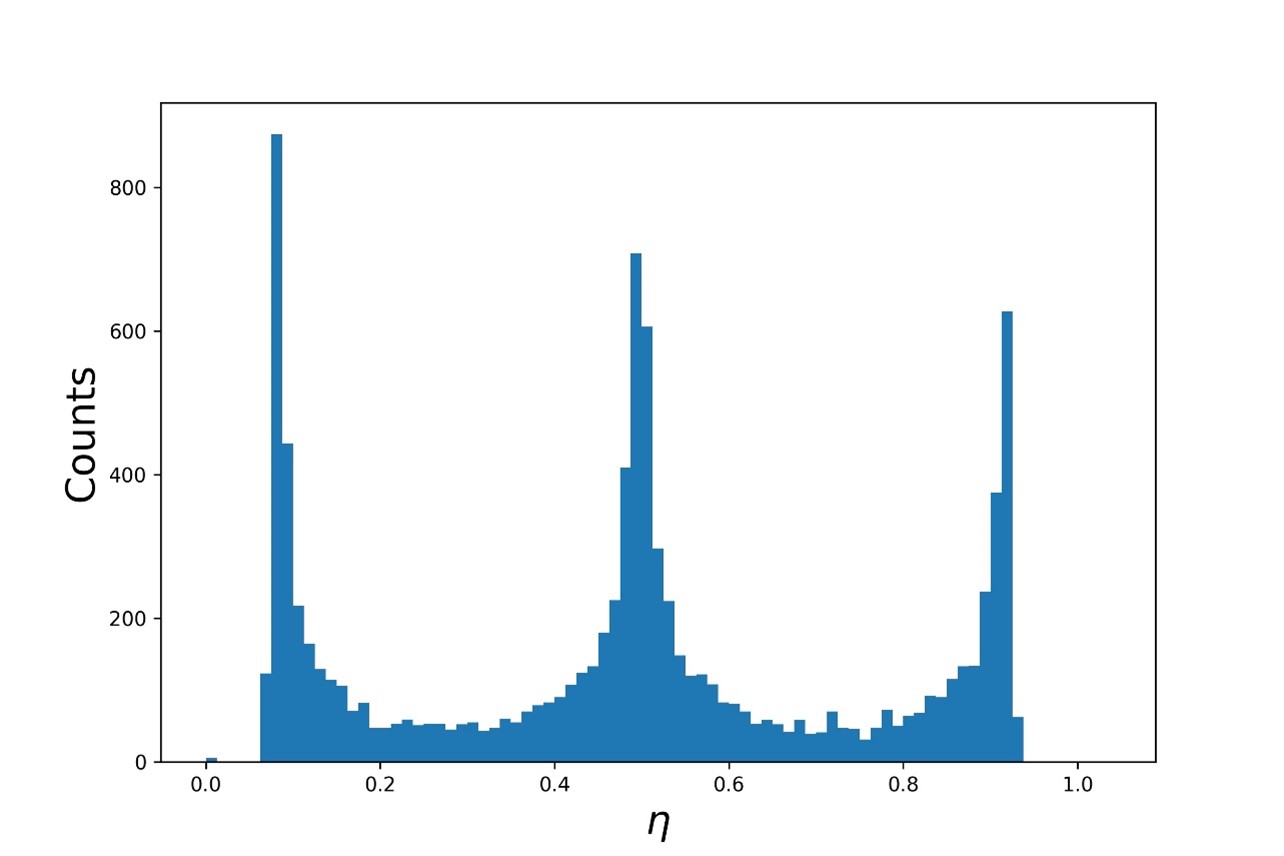}}
\caption{The $\eta$ distribution of the detector set 4 with lithium nuclei incidence.}
\label{fig8}
\end{figure}

\begin{figure}[t]
\centerline{\includegraphics[width=3.5in]{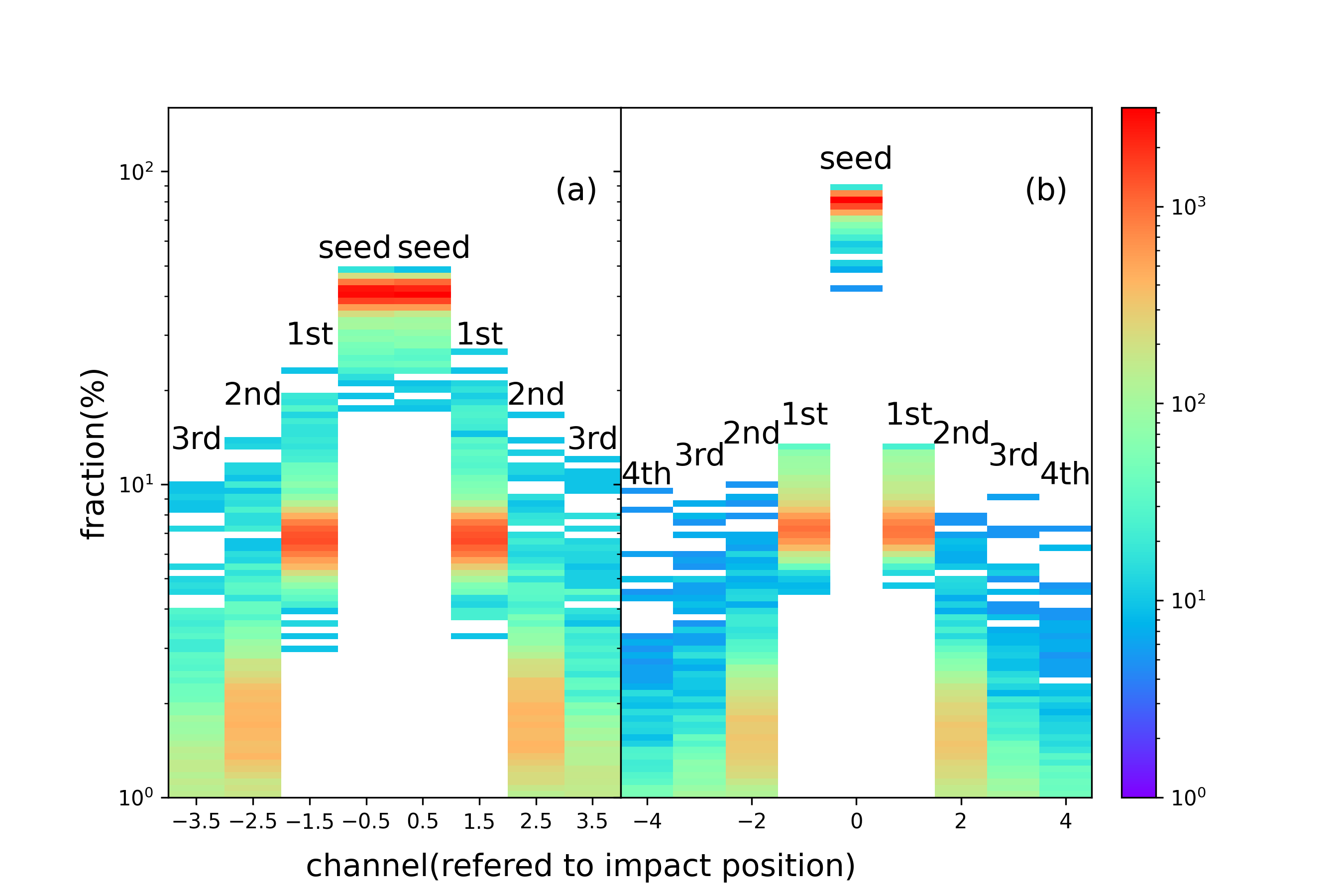}}
\caption{The capacitive coupling fractions of the detector set 4 with lithium nuclei incidence on the floating strips (a) and the readout strips (b). The color refers to the number of entries in each bin.}
\label{fig9}
\end{figure}

\subsection{Capacitive coupling analysis}
\label{subsec:referring}
A typical capacitive coupling effect of readout strip incidence is shown in Fig. 9b. The seed channel collects around 90\% of the total cluster amplitude. The neighboring channels only share a few percent of the total cluster amplitude, and the sharing fraction decreases as the distance increases. The first neighboring channel contributed approximately 7\% of the total cluster amplitude. 
Fig. 9a depicts the capacitive coupling fractions of each channel when lithium nuclei hit the floating strips of detector set 4. The charge collected by the fired floating strip is shared among its neighboring channels, with the sharing fractions decreasing as the distance increases. The fractions of the first neighboring channels amount to approximately 6\% of the total cluster amplitude.
The spectra of the capacitive coupling fractions for the first neighboring channels were accumulated for each ion and each detector set with either readout or floating strip incidences. Each spectrum was fitted with a Gaussian distribution and the Gaussian mean and sigma values were evaluated. A Gaussian fit result of the first neighboring channel of detector set 4 with floating strip incidences is shown in Fig. 10.
Fig. 11a shows the independence of the capacitive coupling fractions of first neighboring channels on the type of light ions with floating strip incidences. This independence can be explained as follows. The capacitive coupling is affected by the incident angle, the charge diffusion during carrier drifting and the capacitor network. During the ion beam test, the SCD prototype detector was installed perpendicular to the beam direction, and the charge diffusion diameters of light ions are negligible with respect to the implant pitch [12]. As a result, the capacitive coupling should only depend on the detector sets. The final capacitive coupling fractions were evaluated as the weighted mean of the capacitive coupling fractions from Z = 1 $\sim$ 7, as shown in Fig. 11 in the dashed lines. It should be noted that the capacitive coupling fractions of ions heavier than nitrogen were not considered due to electronics saturation.  

\begin{figure}[t]
\centerline{\includegraphics[width=3.5in]{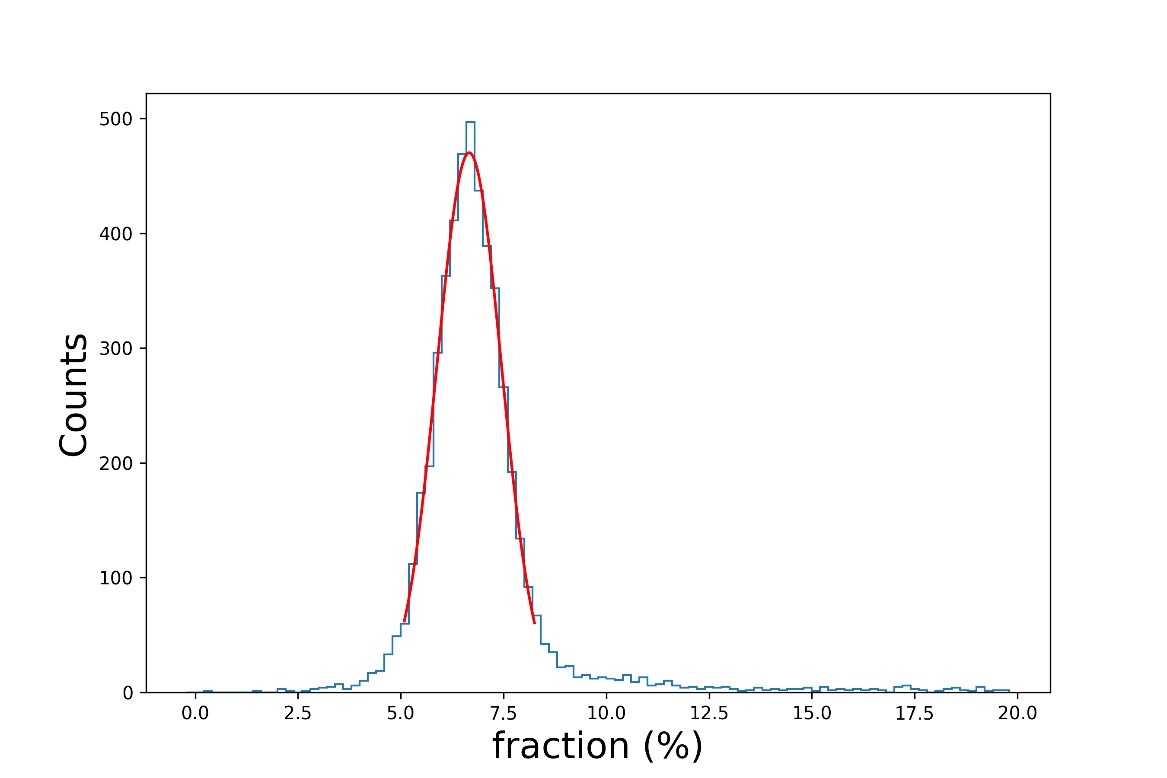}}
\caption{The capacitive coupling fraction distribution of the first neighboring channels for the detector set 4 with lithium nuclei incidence on the floating strips.}
\label{fig10}
\end{figure}

\begin{figure}[t]
\centerline{\includegraphics[width=3.5in]{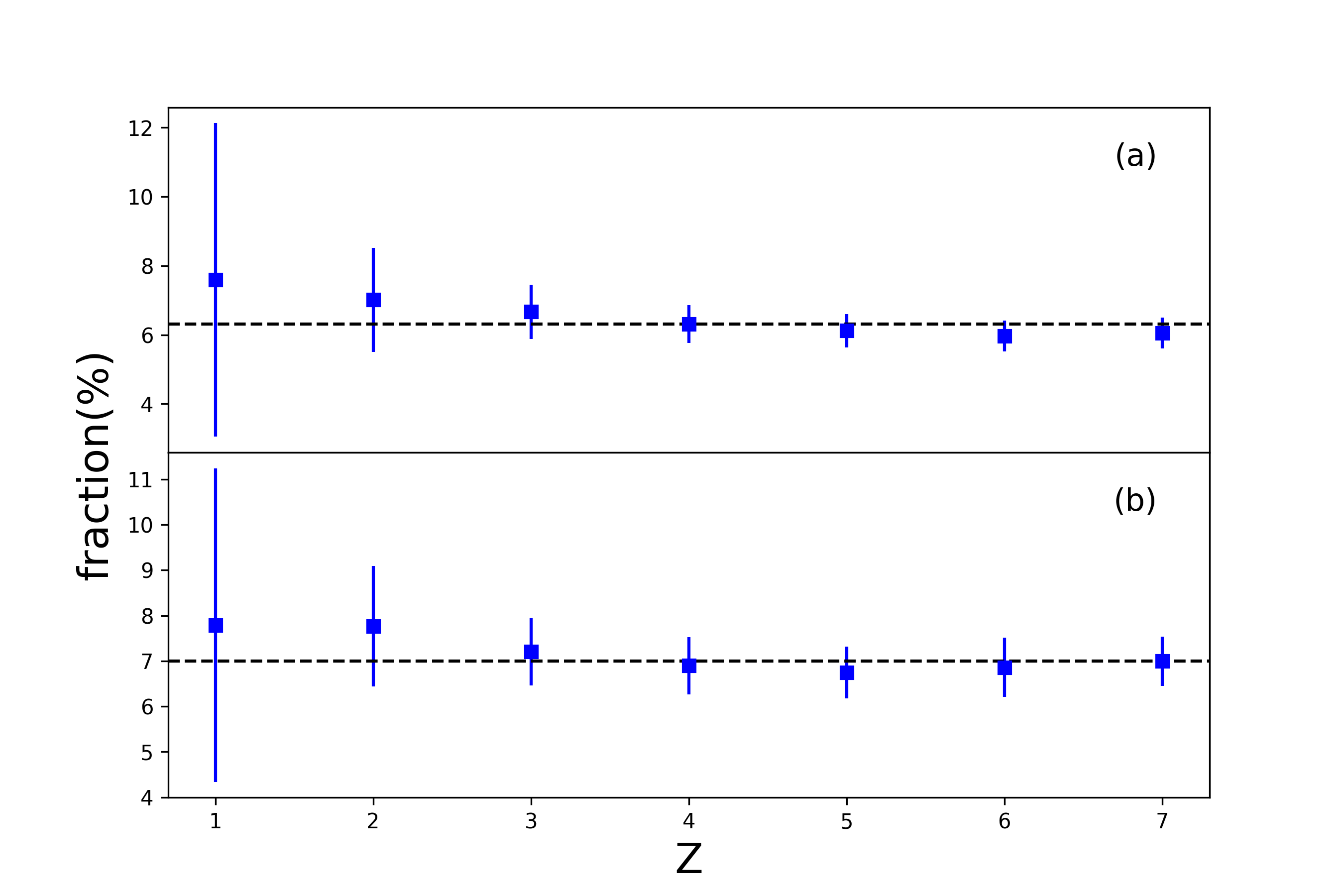}}
\caption{The capacitive coupling fractions of the first neighboring channels with (a) floating strip and (b) readout strip incidences for the detector set 4.}
\label{fig11}
\end{figure}

The same evaluation process of the capacitive coupling fractions of detector set 4 was applied to other detector sets. The relationship between the reciprocal of capacitance values and the capacitive coupling fractions of the first neighboring channels will be discussed in the next section. Fig. 12 shows the relationship between the reciprocal of capacitance values and the capacitive coupling fractions of the first neighboring channels of four detector sets with readout strip and float strip incidences. For both readout and floating strip incidences, a smaller external capacitor increases the capacitive coupling fractions of the first neighboring channels.

\section{SPICE simulation}
In order to obtain the theoretical capacitive coupling fractions, the SCD prototype detector was modelled as a SPICE circuit, as shown in Fig. 13. The second and the third interstrip capacitors were included in the circuit but not shown in Fig. 13 for clear vision. Each readout strip was connected to a preamplifier through an intrinsic coupling capacitor (C\textsubscript{c}) and an external coupling capacitor (C\textsubscript{ce}). The incidence of a charged particle is simulated as a current pulse connected on both sides of a bulk capacitor. The measured capacitance values shown in Table 2 were used in the simulation. The simulation results are shown in Fig. 12 in dashed curves. The SPICE simulation results are consistent with the measured capacitive coupling fractions which depend on the external coupling capacitor.

\begin{figure}[t]
\centerline{\includegraphics[width=3.5in]{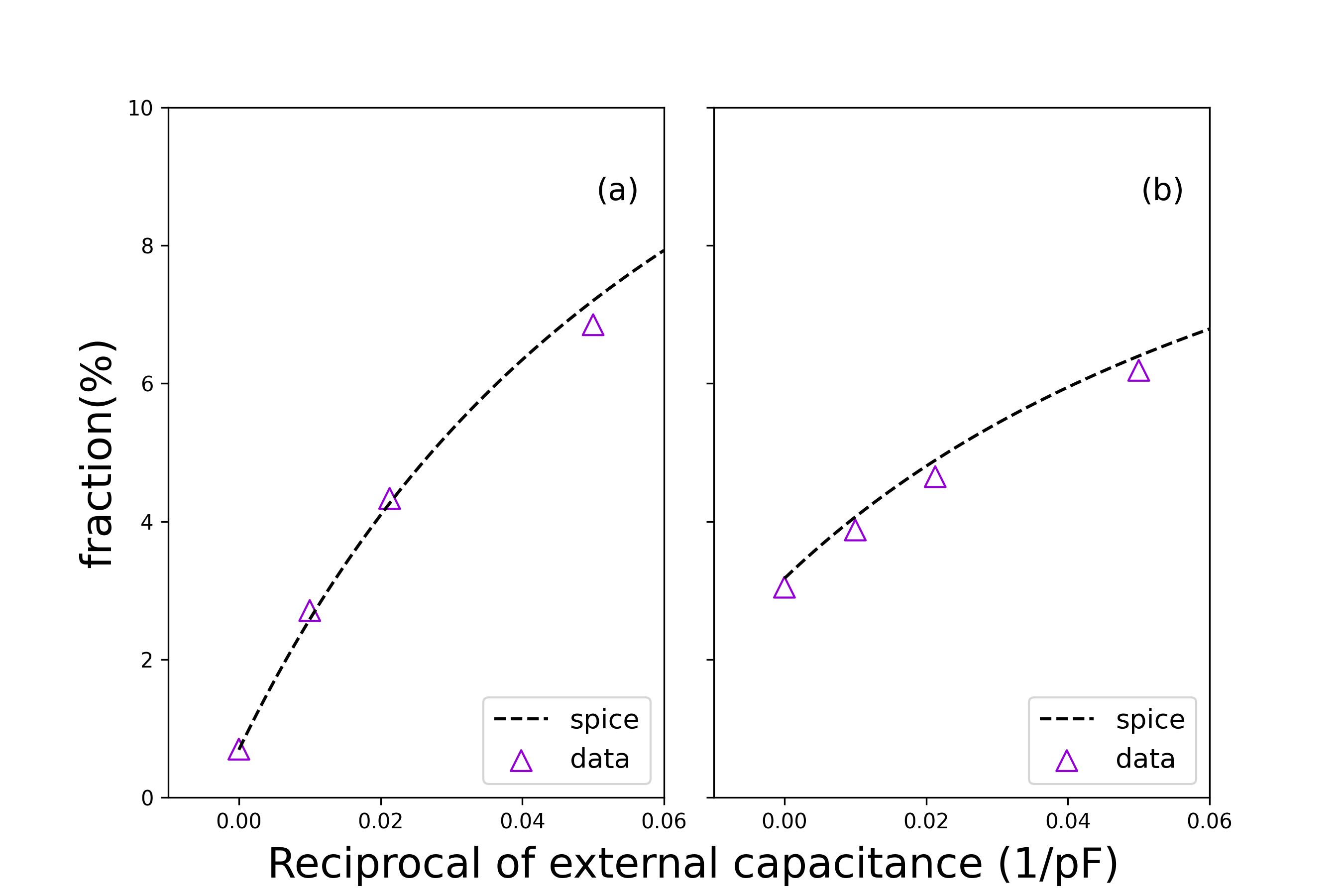}}
\caption{The relationship between the reciprocal of capacitance values and the capacitive coupling fractions of first neighboring channel with readout strip (a) and floating strip (b) incidences in SPICE simulation and in data.}
\label{fig12}
\end{figure}

\begin{figure}[t]
\centerline{\includegraphics[width=3.5in]{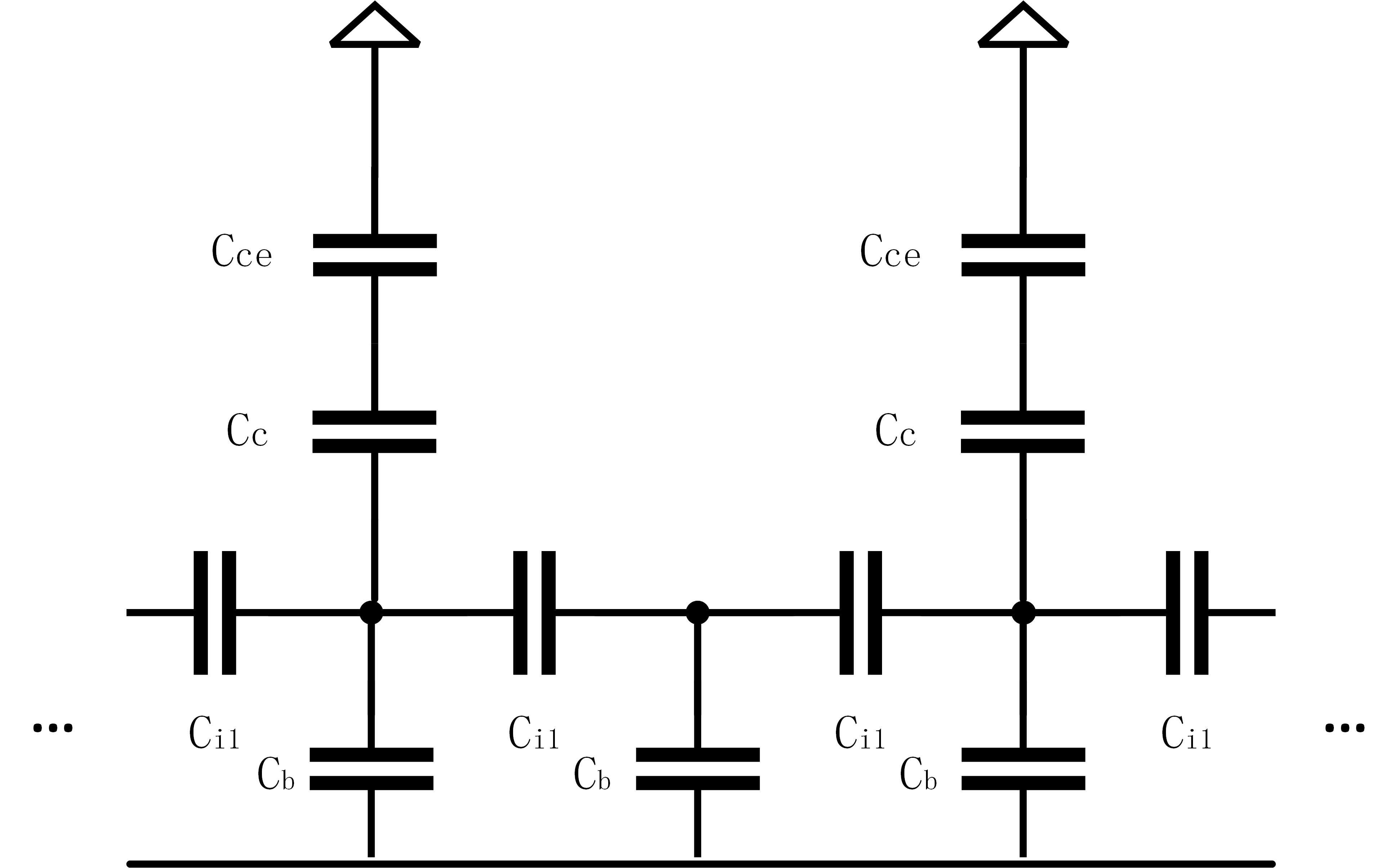}}
\caption{SPICE equivalent model of the microstrip detector.}
\label{fig13}
\end{figure}

\section{Conclusion}
The HERD SCD should have a large dynamic range to measure the charge of Z = 1 $\sim$ 28 cosmic rays. The large dynamic range is proposed to be covered by enhancing the capacitive coupling effect. A HERD SCD prototype detector has been designed to study the capacitive coupling effect dependence of the detector parameters. The detector is divided in several detector sets, and each set has various external capacitors. The detector was illuminated by the ion beams in CERN SPS and the capacitive coupling fractions with readout strip and floating strip incidences were evaluated. The capacitive coupling fractions were not sensitive to the species of ions, but were dependent on the impact position. As the distance to the impact position increases, the capacitive coupling fractions decrease. In addition, the detector parameters and the external coupling capacitors also affect the capacitive coupling fractions. A SPICE simulation has been implemented by modelling the SCD detector as a capacitor network, and the simulation results were consistent with the measurements. The knowledge of the capacitive coupling effect dependence to the detector parameters helps to optimize the design of the SCD detector.



\section*{References}

\def\refname{\vadjust{\vspace*{-1em}}} 


\begin{thebibliography}{00}
	\bibitem{1} S. N. Zhang, O. Adriani, S. Albergo, G. Ambrosi, Q. An, T. W. Bao, ~\textit{et al.}, ``The high energy cosmic-radiation detection (HERD) facility onboard China’s Space Station," in {\textit {Space Telescopes and Instrumentation 2014: Ultraviolet to Gamma Ray}}, Canada: Montréal, 2014.
	\bibitem{2}	J. R. Hörandel, ``Models of the knee in the energy spectrum of cosmic rays," {\textit {Astropart. Phys.}}, vol. 21, no. 3, pp.  241-265, 2004, DOI. 10.1016/j.astropartphys.2004.01.004.
\bibitem{3}		Y. W. Dong, Z. Quan, J. J. Wang, M. Xu, S. Albergo, F. Ambroglini, ~\textit{et al.}, ``Experimental verification of the HERD prototype at CERN SPS," in {\textit {Space Telescopes and Instrumentation 2016}}, UK: Ultraviolet to Gamma Ray, Edinburgh, 2016.
	\bibitem{4}	D. Kyratzis, `` Latest advancements of the HERD space mission," {\textit {Nucl. Instrum. Methods Phys. Res. A}}, vol. 1048, pp.  167970, 2023, DOI.  10.1016/j.nima.2022.167970.
\bibitem{5}	R. Qiao, W. X. Peng, X. Z. Cui, G. Q. Dai, Y. F. Dong, R. R. Fan, ~\textit{et al.}, ``A charge sharing study of silicon microstrip detectors with electrical characterization and SPICE simulation," {\textit {Adv. Space Res.}}, vol. 64, no. 12, pp. 2627-2633, 2019, DOI.  10.1016/j.asr.2019.07.005.
\bibitem{6}	M. A. Frautschi, M. R. Hoeferkamp and S.C. Seidel, ``Capacitance measurements of double-sided silicon microstrip detectors," {\textit {Nucl. Instrum. Methods Phys. Res. A}}, vol. 378, no. 1-2, pp.  284-296, 1996, DOI. 10.1016/0168-9002(96)00467-6. 
\bibitem{7}Micron. (2021). Micron strip detectors/ single sided [Online]. Available: https://www.micronsemiconductor.co.uk/
\bibitem{8}Ideas. (2023). IDE1140 [Online]. Available: https://ideas.no/
\bibitem{9}M. Krammer and H. Pernegger, `` Signal collection and position reconstruction of silicon strip detectors with 200 $\mu$m readout pitch," {\textit {Nucl. Instrum. Methods Phys. Res. A}}, vol. 397, no. 2-3, pp. 232-242, 1997, DOI.  10.1016/S0168-9002(97)00802-4.
\bibitem{10}M. Prest, G. Barbiellini, G. Bordignon, G. Fedel, F. Liello, F. Longo, ~\textit{et al.}, ``The AGILE silicon tracker: an innovative $\gamma$-ray instrument for space," {\textit {Nucl. Instrum. Methods Phys. Res. A}}, vol. 501, no. 1, pp. 280-287, 2003, DOI. 10.1016/S0168-9002(02)02047-8.
\bibitem{11}Y. F. Dong, F. Zhang, R. Qiao, W. X. Peng, R. R. Fan, K. Gong, ~\textit{et al.}, ``DAMPE silicon tracker on-board data compression algorithm," {\textit {Chinese Phys. C}}, vol. 39, no. 11, pp. 116202, 2015, DOI. 10.1088/1674-1137/39/11/116202.
\bibitem{12}E. Belau, R. Klanner, G. Lutz, E. Neugebauer, H. J. Seebrunner and A. Wylie, ``Charge Collection in silicon strip detectors," {\textit {Nucl. Instrum. Methods Phys. Res. A}}, vol. 214, no. 2, pp. 253-260, 1983, DOI. 10.1016/0167-5087(83)90591-4.
\end{thebibliography}
\end{document}